\documentclass{webofc}

\usepackage[varg]{txfonts}   
\usepackage{hyperref}
\usepackage{url}
\usepackage{xcolor}
\hypersetup{colorlinks=true,citecolor=blue,urlcolor=blue,linkcolor=blue}

\begin{document}

\title{From Commissioning to Precision Data-Taking: Resolving Operational Challenges in the Nab Detector Systems}

\author{
\firstname{L. J. }\lastname{Broussard}\inst{1}\fnsep\thanks{\email{broussardlj@ornl.gov}} 
\and
\firstname{H.} \lastname{Acharya}\inst{2}
\and
\firstname{R.} \lastname{Alarcon}\inst{3}
\and
\firstname{S.} \lastname{Bae{\ss}ler}\inst{4}\fnsep\inst{1}  
\and
\firstname{M.} \lastname{Benoit}\inst{1}
\and
\firstname{K.} \lastname{Borah}\inst{2}
\and
\firstname{C. L.} \lastname{Britton}\inst{1}
\and
\firstname{E.} \lastname{Brown}\inst{5}
\and
\firstname{J.} \lastname{Choi}\inst{6}\fnsep\inst{7} 
\and
\firstname{S.} \lastname{Clymer}\inst{3}
\and
\firstname{C.} \lastname{Crawford}\inst{2}
\and
\firstname{N.} \lastname{Ericson}\inst{1}
\and
\firstname{L.} \lastname{Fabris}\inst{1}
\and
\firstname{N.} \lastname{Fomin}\inst{5}
\and
\firstname{J.} \lastname{Fry}\inst{8}
\and
\firstname{R.} \lastname{Godri}\inst{5} 
\and
\firstname{F. M.} \lastname{Gonzalez}\inst{1} 
\and
\firstname{A.} \lastname{Hagemeier}\inst{4}
\and
\firstname{J.} \lastname{Hamblen}\inst{9}
\and
\firstname{S.} \lastname{Hollander}\inst{1} 
\and
\firstname{A.} \lastname{Jezghani}\inst{10}  
\and
\firstname{K.} \lastname{Leung}\inst{11}  
\and
\firstname{N.} \lastname{Macsai}\inst{13}  
\and
\firstname{M.} \lastname{Makela}\inst{12} 
\and
\firstname{D.} \lastname{Mathews}\inst{1} 
\and
\firstname{P. L. } \lastname{McGaughey}\inst{12}  
\and
\firstname{A.} \lastname{Mendelsohn}\inst{13}  
\and
\firstname{J.} \lastname{Mirabal}\inst{12}  
\and
\firstname{P. E.} \lastname{Mueller}\inst{1}
\and
\firstname{A.} \lastname{Nelsen}\inst{2}
\and
\firstname{S. I.} \lastname{Penttil\"{a}}\inst{1}
\and
\firstname{D.} \lastname{Po\v{c}ani\'c}\inst{4} 
\and
\firstname{J. C.} \lastname{Ramsey}\inst{1}
\and
\firstname{K.} \lastname{Reed}\inst{1}
\and
\firstname{L.} \lastname{Richburg}\inst{5}  
\and
\firstname{A.} \lastname{Saunders}\inst{1}
\and
\firstname{W.} \lastname{Schreyer}\inst{1}
\and
\firstname{A.} \lastname{Shelby}\inst{6}\fnsep\inst{7}
\and
\firstname{A. R.} \lastname{Young}\inst{6}\fnsep\inst{7}
}

\institute{Oak Ridge National Laboratory, Oak Ridge, TN 37831, USA
\and
           University of Kentucky, Lexington, KY 40506, USA
\and
           Arizona State University, Tempe, AZ 85287, USA 
\and
           University of Virginia, Charlottesville, VA 22904, USA
\and
           University of Tennessee, Knoxville, TN 37996, USA
\and
           North Carolina State University, Raleigh, NC 27695, USA
\and
           Triangle Universities Laboratory, Duke University, Durham, NC 27708, USA
\and
           Eastern Kentucky University, Richmond, KY 40475, USA
\and
           University of Tennessee at Chattanooga, Chattanooga, TN 37403, USA
\and
           Partnership for an Advanced Computing Environment, Georgia Tech, Atlanta, GA 30332, USA
\and
           Montclair State University, Montclair, NJ 07043, USA
\and
           Los Alamos National Laboratory, Los Alamos, NM 87545, USA
\and
           University of Manitoba, Winnipeg, MB R3T 2N2, Canada
          }

\abstract{ 
Our understanding of the weak mixing of quarks, described by the Cabibbo Kobayashi Maskawa (CKM) matrix, currently presents an anomaly. 
Thanks to major strides in both theory and experiment, improved precision in determinations of the first row of matrix elements has revealed disagreement with the expectation of unitarity. The Nab experiment at the Spallation Neutron Source is designed to precisely extract the first matrix element $V_{ud}$ and shed light on experimental tensions within the neutron beta decay dataset. Nab’s asymmetric spectrometer allows coincident reconstruction of the decay proton and electron energies, which will be used to determine the electron-neutrino correlation coefficient, and thus (with the neutron lifetime) determine $V_{ud}$. This unique approach has provided a more comprehensive view of neutron beta decay, including a first observation of the full momentum phase space of the decay above detector thresholds and limits on exotic neutron states. Recent upgrades to the Nab detector system have improved the robustness and stability of the detector performance in terms of proton detection efficiency, noise performance, and detector segment availability, setting the stage for high precision physics data-taking.} 

\maketitle

\section{Introduction}
\label{intro}

The standard model's description of the mixing of quarks in the weak interaction must obey a simple yet powerful principle---it must be unitary. The mixing of quarks is represented by the Cabibbo-Kobayashi-Maskawa (CKM) matrix, and the most precise test of its unitarity relies on the first two elements: $V_{ud}$, which describes up-down quark transitions in nuclear systems, and $V_{us}$, primarily determined from kaon decays.  Currently there is a discrepancy with unitarity, which now stands in tension by a few standard deviations~\cite{ParticleDataGroup:2024cfk}. Internal inconsistencies are observed within the systems used to determine $V_{ud}$ and $V_{us}$ and must be resolved before the unitarity test can be applied. 

The achievable precision of $V_{ud}$ determined using the set of superallowed Fermi nuclear decays is limited by uncertainty in the nuclear structure corrections, which may be challenging to improve~\cite{Hayen:2024xjf}. The neutron system offers a compelling opportunity to inform the situation in the near term, via a determination of both the lifetime plus the ratio $\lambda = g_\text{A}/g_\text{V}$ of axial-vector ($g_\text{A}$) and vector ($g_\text{V}$) couplings, if experimental inconsistencies within those datasets can be addressed. There is a long-standing discrepancy between different techniques for measuring the neutron lifetime~\cite{ParticleDataGroup:2024cfk}. The most precise determinations of $\lambda=-1.27641\pm0.00056$~\cite{Markisch:2018ndu} from the beta-asymmetry and $\lambda=-1.2668\pm0.0027$~\cite{Beck:2023hnt} from the electron-neutrino correlation disagree by about 3\,$\sigma$. The Nab experiment is pursuing a new measurement of the electron-neutrino correlation with goal sensitivity of $\Delta \lambda/|\lambda|=0.04\%$. This result would lay the groundwork for the neutron to provide an independent and precise determination of $V_{ud}$ without the uncertainties of nuclear corrections, and shed light on hints of interactions not in the standard model. 

\section{Experimental Approach}

Nab uses a powerful experimental approach that allows for the reconstruction of the full momentum phase space of neutron beta decay accessible above detection thresholds, from which the electron-neutrino correlation can be extracted~\cite{Nab:2008cwh,Nab:2012hnc, Fry:2018kvq}. The novel asymmetric magnetic spectrometer captures charged beta decay products from the nominally unpolarized neutron beam to one of two detector systems.  The magnet profile ensures upward-going protons are nearly parallel to the magnetic field, which enables an accurate determination of the proton's energy from its observed time of flight relative to the electron. 

The Nab experiment began collecting commissioning data in 2023 with results reported in~\cite{Nab:2025tgs}. This first data-taking campaign had the primary goal to ensure operation of the experimental subsystems and understand the optimal data-taking conditions. The measured phase space of the decay from this dataset provided a clear demonstration of the working principles of the spectrometer. The tight kinematic constraints available in Nab also enabled new experimental limits on a hypothesized excited state of the neutron, proposed as an explanation of the long-standing discrepancy observed between neutron lifetime measurements made with a cold neutron beam versus trapped ultracold neutrons~\cite{Koch:2024cfy}.

\section{Commissioning Challenges}

The commissioning data-taking demonstrated the robustness of the general experimental approach, but also revealed challenges which endangered the ultimate physics goals of the experiment. As a result of the data-taking campaign, three issues were identified as priorities.
\begin{enumerate}
\item Protons were detected at $\sim$10\,keV instead of $\sim$20\,keV as expected.
\item The detector electronics became unstable when fully powered.
\item Loss of connections in the electronics resulted in many unreporting detector segments.
\end{enumerate}
We report here on the outcome of improvements to the detector system and associated operations and impact for Nab physics data-taking. Later datasets also benefited from a corrected timing synchronization routine, utilization of the full-scale calibration system, and greater stability of data-taking conditions, which are not described in this work.

\subsection{Unexpected Reduction of Proton Energy}

\begin{figure}[h]
\centering
\includegraphics[width=\textwidth,clip]{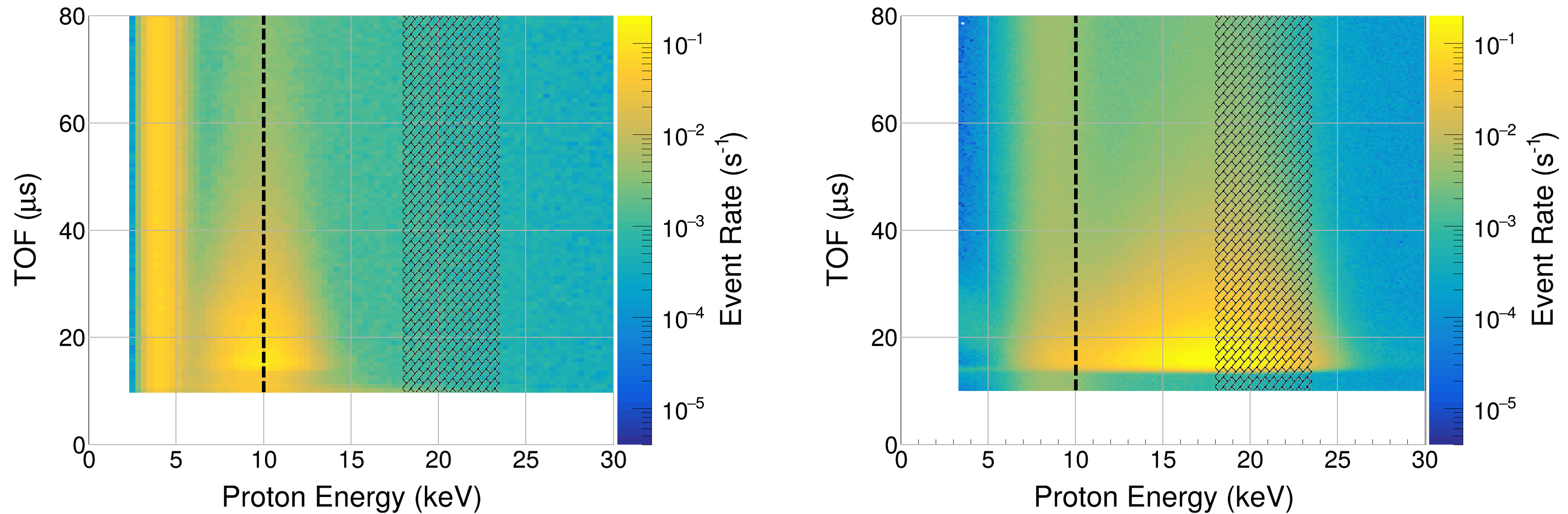}
\caption{Detected events as a function of proton Time Of Flight (TOF) and proton energy during commissioning in 2023 (left) and after improvements to detector qualification procedures in 2024 (right), assuming a rough energy calibration, and mean proton energy as observed in 2023 (left line) and the range of means expected based on 50--100\,nm deadlayer (right hatch).}
\label{fig-1}  
\end{figure}

In Nab, protons are expected to deposit roughly 20\,keV in the active region of the detector, with few keV variation depending on detector deadlayer~\cite{Salas-Bacci:2014xpo}. While accurate energy calibration of protons is not needed, efficient detection of protons is required to avoid systematic bias in the electron-neutrino correlation~\cite{Fry:2018kvq}. The detector used for proton detection was originally characterized using a proton beam facility~\cite{Manitoba:2025} which confirmed the detector response matched expectations. However, during the 2023 data-taking at ORNL, an unexpectedly low proton energy of $\sim$10\,keV was observed (Figure~\ref{fig-1} left). Several possible causes for lost proton energy were investigated, including temporary formation of ice layers, deposits from contamination, deposits or detector damage from high voltage breakdown, and mis-calibration. As a result of these investigations, more robust detector qualification procedures were implemented, detector test environments were improved to achieve ultrahigh vacuum, and provision for silicon ``witness'' samples was added for more detailed surface analyses. In data-collection campaigns in 2024 and 2025, use of pristine detectors under these procedures resulted in protons having detectable energy within the expected range (Figure~\ref{fig-1} right).

\subsection{Detector Electronics Instability}

\begin{figure}[h]
\centering
\includegraphics[width=\textwidth,clip]{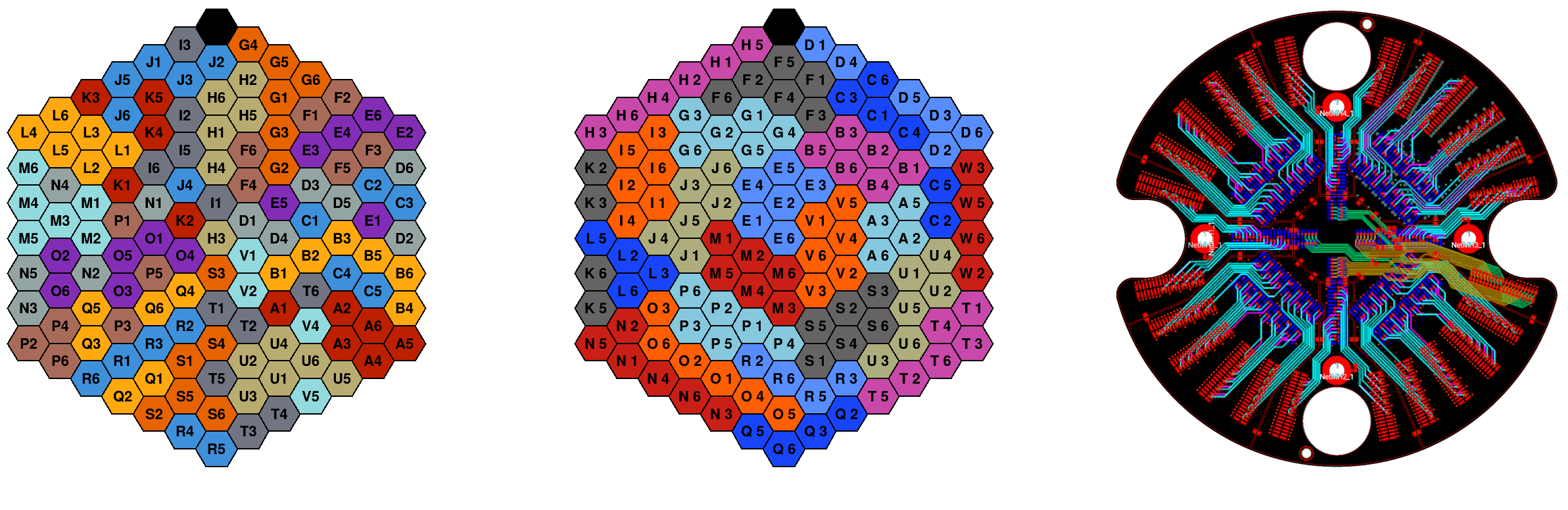}
\caption{(Left) Mapping of preamplifier circuit board to detector segment during the 2023 commissioning dataset and (center) the redesigned mapping used in 2024 and later. Colors separate different boards. Hexagonal segments are enumerated by board identifier letter plus channel number.  (Right) Updated layout of the 6.54\,in diameter adapter board which maps the 6\,channel circuit boards and thermometers (23 + 1 outer red connectors) to a vacuum feedthrough (12 inner blue connectors). Different colored traces reside on different board layers. (color online)}
\label{fig-2} 
\end{figure}

The electronics circuit used in the Nab experiment has been demonstrated to meet the performance requirements for Nab~\cite{Broussard:2016gqg}. However, when scaled to the fully instrumented 127 channel system, the entire system became unstable and could not be operated with all channels powered. In testing at the Manitoba proton facility and during the 2023 commissioning campaign, the problem was mitigated simply by shutting off power to some of the preamplifier circuit boards. The original design mapped these 6 channel boards to a roughly radial slice of the detector (Figure~\ref{fig-2} left), evenly distributing signal rate across boards. A consequence was that unpowering any board removed some central pixels which primarily see beta decay events. The new mapping instead allows boards containing only outer ring pixels that detect backgrounds to be unpowered (Figure~\ref{fig-2} center). The layout of the most complicated adapter board is shown in Figure~\ref{fig-2} (right).

Additional modifications were implemented which enhanced the overall stability of the full-scale system.  The total power draw of the preamplifiers was reduced by about 1/3 by removing the optional shaping circuit. The redesigned mapping in the adapter boards reduced the total trace length of many channels and reduced cross-talk. New circuit boards were fabricated including a slightly thicker ground layer. These changes effectively eliminated the rail-to-rail oscillations of the electronics during data collection campaigns after commissioning. Further refinements to the power delivery and ground planes intended to boost stability and reduce noise have shown promising results in bench testing.

\subsection{Loss of detector segments}

\begin{figure}[h]
\centering
\includegraphics[width=12cm,clip]{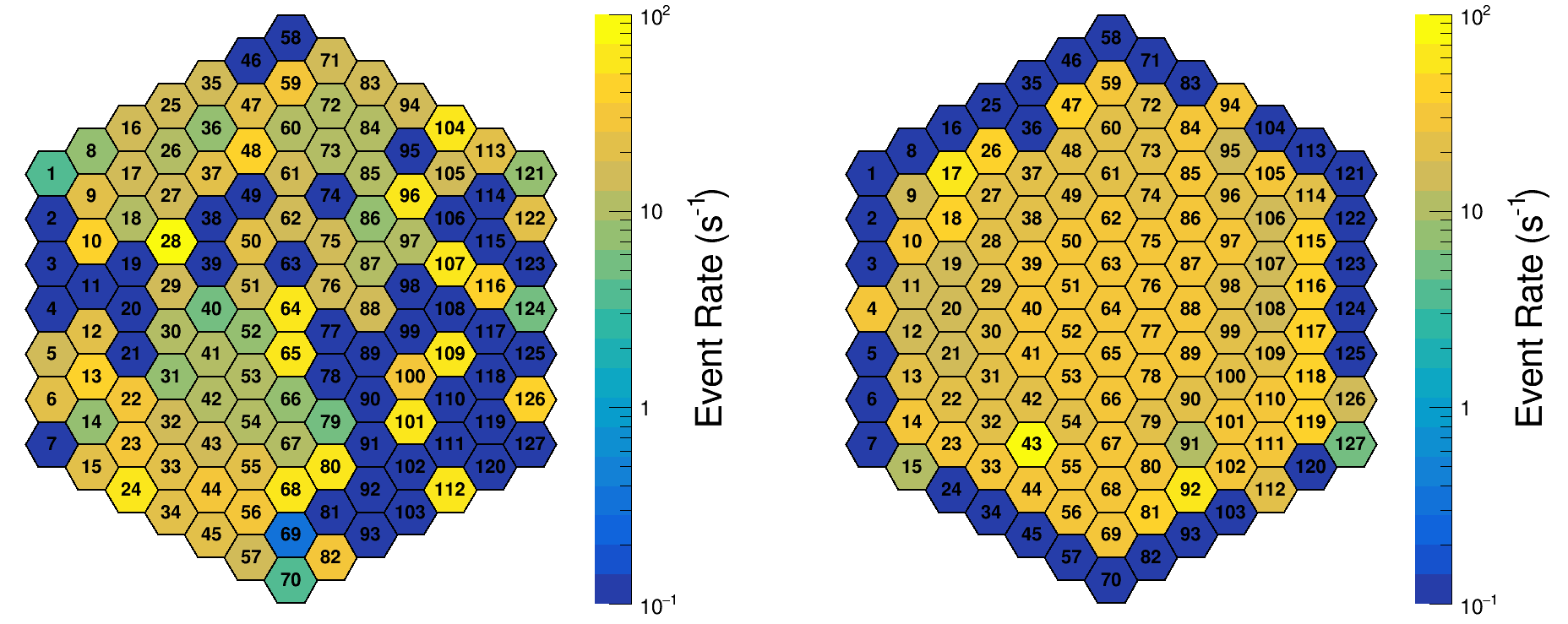}
\caption{The detected total trigger rate (color scale) for detector segments for typical runs during the 2023 commissioning data-collection (left) and during the 2025 data-collection (right). Darker colors indicate segments which are non-reporting. Noise and threshold conditions varied among pixels and datasets, in particular in 2023.}
\label{fig-3} 
\end{figure}

\begin{figure}[h]
\centering
\includegraphics[width=\textwidth,clip]{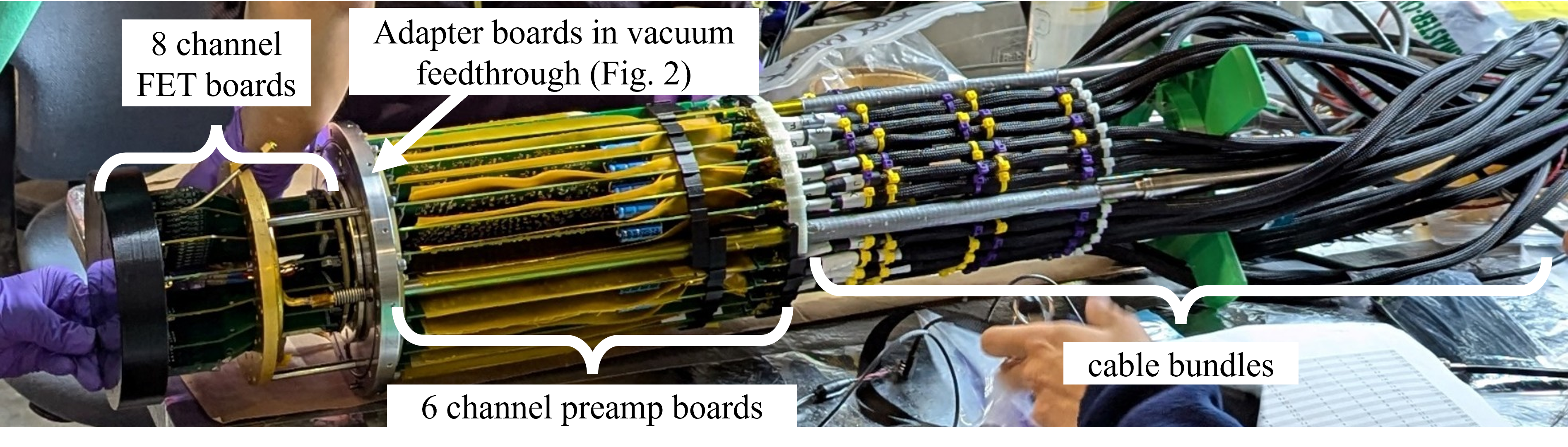}
\caption{Updated electronics package including 8 channel circuit boards with Field Effect Transistors (FET), 6 channel boards with amplification stages, adapter boards, and cabling.}
\label{fig-4}  
\end{figure}

In Nab, electron energy cannot be properly reconstructed without accounting for backscattering events which may appear in neighboring detector segments. Thus, loss of any segment dramatically reduces the effectiveness of the neighbors as well. Loss of reporting segments during installation procedures proved to be a persistent challenge through the 2023 commissioning data collection (Figure~\ref{fig-3} left), primarily due to connector damage, accidental disconnects, and unpowering of preamplifier boards required for stability. After the 2023 commissioning data-collection, a series of modifications were implemented to reduce risk during the installation procedure and increase the robustness of the electronics package (Figure~\ref{fig-4}). The 2024 and 2025 data-collection campaigns saw significant recovery of the number of operational central pixels (Figure~\ref{fig-3} right).

\section{Summary}

The first data-collection campaign in Nab has provided first physics results and demonstrated the principles of operation of the experiment. Thanks to a series of upgrades relevant to the detection system, three critical areas have been addressed. In the physics data-taking campaigns of 2024 and 2025, we have maintained operation of the most important central pixels, eliminated noise instability, and demonstrated the expected detector response to protons. Thanks to these and other operational improvements, the Nab experiment is well positioned to deliver much needed input on the tension observed in determinations of $\lambda$ from the neutron dataset. This will enable a robust and precise determination of the CKM parameter $V_{ud}$, used in one of the most precise tests of our understanding of the weak mixing of quarks.

\section{Acknowledgements}

This work was supported through the US Department of Energy (DOE) (Contracts DE-AC05-00OR22725, 
DE-SC0008107, DE-SC0014622, 
89233218CNA000001 under proposal LANLEEDM, 
DE-SC0019309, 
DE-FG02-97ER41042, 
DE-FG02-03ER41258), 
and through the National Science Foundation (NSF) (Contracts 
PHY-0855584, 
PHY-2412782, 
PHY-2111363, 
PHY-1126683, 
PHY-2232117, 
PHY-2209590, 
PHY-2213411, and PHY-2412846), 
by the Natural Sciences and Engineering Research Council of Canada (NSERC) (Contracts NSERC-SAPPJ-2019-00043 and NSERC-SAPPJ-2022-00024), 
and in part by the US Department of Energy, Office of Science, Office of Workforce Development for Teachers and Scientists (WDTS) Graduate Student Research (SCGSR) program. 
This research was supported through research cyberinfrastructure resources and services provided by the Partnership for an Advanced Computing Environment (PACE) at the Georgia Institute of Technology, Atlanta, Georgia, USA. 
This research used resources at the Spallation Neutron Source, a DOE Office of Science User Facility operated by the Oak Ridge National Laboratory.

\bibliography{main.bib}

\end{document}